\documentclass[reprint, superscriptaddress, amsmath, amssymb, aps, prb]{revtex4-2}
\usepackage[dvipdfmx]{graphicx}
\usepackage{dcolumn}
\usepackage{bm}
\usepackage[dvipdfmx]{hyperref}
\usepackage{comment}
\usepackage{color}
\usepackage{physics}

\begin{document}
\preprint{APS/123-QED}
\title{Exceptionally large winding number of a finite-size topological superconductor}
\author{Satoshi Ikegaya}
\affiliation{Institute for Advanced Research, Nagoya University, Nagoya 464-8601, Japan}
\affiliation{Department of Applied Physics, Nagoya University, Nagoya 464-8603, Japan}
\author{Shingo Kobayashi}
\affiliation{RIKEN Center for Emergent Matter Science, Wako, Saitama, 351-0198, Japan}
\author{Yasuhiro Asano}
\affiliation{Department of Applied Physics, Hokkaido University, Sapporo 060-8628, Japan}
\date{\today}
%

\begin{abstract}
We study finite-size-induced topological phenomena in unconventional superconductors.
Specifically, we focus on a thin film with a persistent spin texture, fabricated on a high-$T_{\text{c}}$ cuprate $d_{xy}$-wave superconductors.
In two-dimensional $d_{xy}$-wave superconductors, flat-band Andreev bound states appear at the edges.
As the system narrows, these bound states acquire an energy gap due to finite-size hybridization and spin-orbit coupling of the persistent spin texture.
This induced gap gives rise to the emergence of a topological phase, characterized by an exceptionally large one-dimensional winding number that scales with the film width.
We demonstrate the appearance of highly degenerate zero-energy states, leading to anomalous perfect charge transport in dirty superconducting junctions.
These findings provide a promising platform for exploring fascinating topological superconducting phases driven by gapped Andreev bound states.
\end{abstract}
\maketitle

\textit{Introduction and outline.}
Topological phases of matter, characterized by nontrivial topological invariants associated with gapped band structures, have been a central focus of condensed matter physics.
A hallmark of these phases, as dictated by the bulk-boundary correspondence, is the emergence of exotic boundary states. 
The intrinsic stability of these states holds promise for future applications in quantum technologies~\cite{nayak_08,kane_10,zhang_11,das_16,sato_17}.
The topological invariant of gapped band structures is determined primarily by two factors: the dimensionality and the symmetry of the systems~\cite{schnyder_08}.   
The relationship between topological properties and dimensionality is of particular importance in quasi-$(d-1)$-dimensional systems,
where the size of a $d$-dimensional system is reduced along one direction.
In such systems, topological properties can undergo marked changes as the system approaches the dimensional crossover regime~\cite{Xue_10,lee_10}.

A representative example is found in quasi-two-dimensional thin films of topological insulators.
In three-dimensional topological insulators, Dirac surface states appear as a consequence of a $\mathbb{Z}_2$ topological invariant.
In quasi-two-dimensional systems, finite-size effects induce hybridization between the Dirac surface states on the top and bottom surfaces, leading to two-dimensional (2D) topological insulators~\cite{Xue_10,Hasegawa_10,Ando_12,Vikram_22}, depending on the parity of the number of stacked 2D layers~\cite{Linder_09,Shou-Cheng_10}. 
These coupled Dirac surface states can be further engineered by additional perturbations.
For instance, applying an exchange potential opens a gap in the Dirac spectrum and stabilizes a topological phase characterized by the Chern number~\cite{fang_10}, 
while an extended $s$-wave pairing potential can induce a $\mathbb{Z}_2$ topological superconducting phase~\cite{mele_13,sarma_21}.

Despite the extensive study of topological phases originating from Dirac surface states%
~\cite{kane_08,fang_10,Shen_10(1),Shen_10(2),Tanaka_12,Tanaka_14,nagaosa_14,zhang_15,Ilya_18,hu_20,sarma_21,cook_23,cook_23(2),cook_25},  
topological phases driven by Andreev bound states (ABSs) of superconductors (SCs) remain poorly understood.
A known case involves Caroli-de Gennes-Matricon bound states in vortex cores%
~\cite{Hosur_11,bernevig_14,Shou-Cheng_16,hu_19,Coleman_19,furusaki_20,Huang_20,Hosur_21,zhang_22,Kobayashi_23,Rui-Xing_23},
which result from the phase winding around a vortex rather than a finite-size effect.
Therefore, superconducting analogs of topological phases induced by finite-size effects remain largely unexplored.
\begin{figure*}[t]
\begin{center}
\includegraphics[width=1\textwidth]{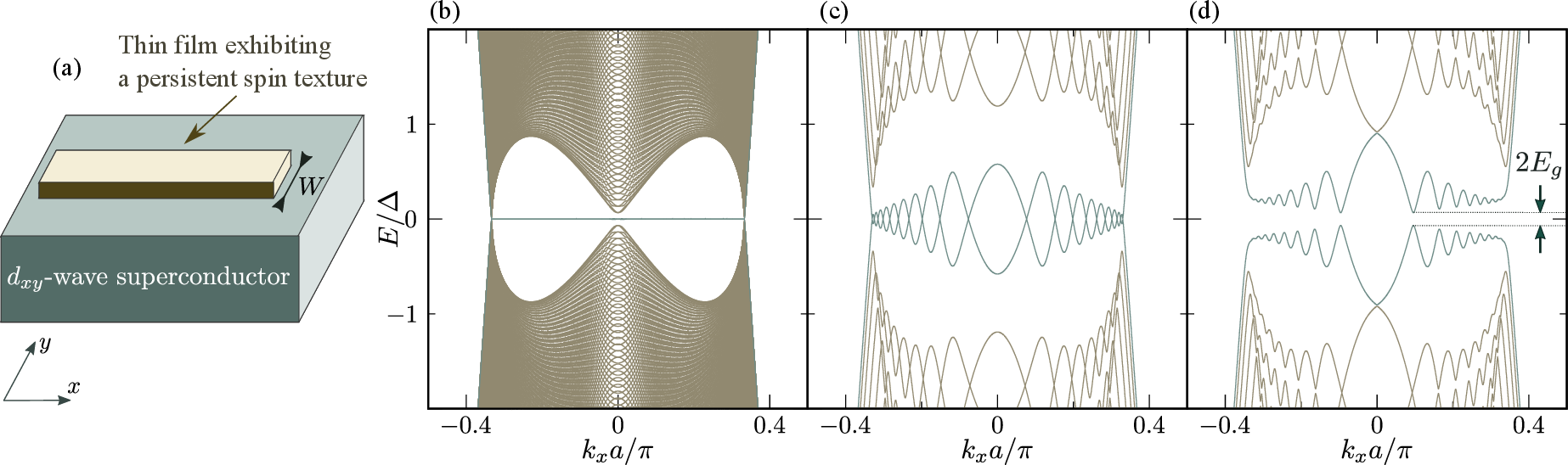}
\caption{(a) Schematic of the system under consideration, where the unidirectional SOC potential of the PST and the proximity-induced $d_{xy}$-wave pair potential coexist in the thin film.
(b)--(c) Energy spectra as a function of $k_x$ for (b) $(W,\lambda)=(800a,0)$, (c) $(W,\lambda)=(30a,0)$, and (d) $(W,\lambda)=(30a,0.5t)$, with an open boundary condition applied in the $y$ direction.
As shown in (d), the energy spectrum of the ABS exhibits a gap, $E_g$, resulting from the finite-size hybridization and unidirectional SOC.
}
\label{fig:figure1}
\end{center}
\end{figure*}

\begin{figure}[t]
\begin{center}
\includegraphics[width=0.375\textwidth]{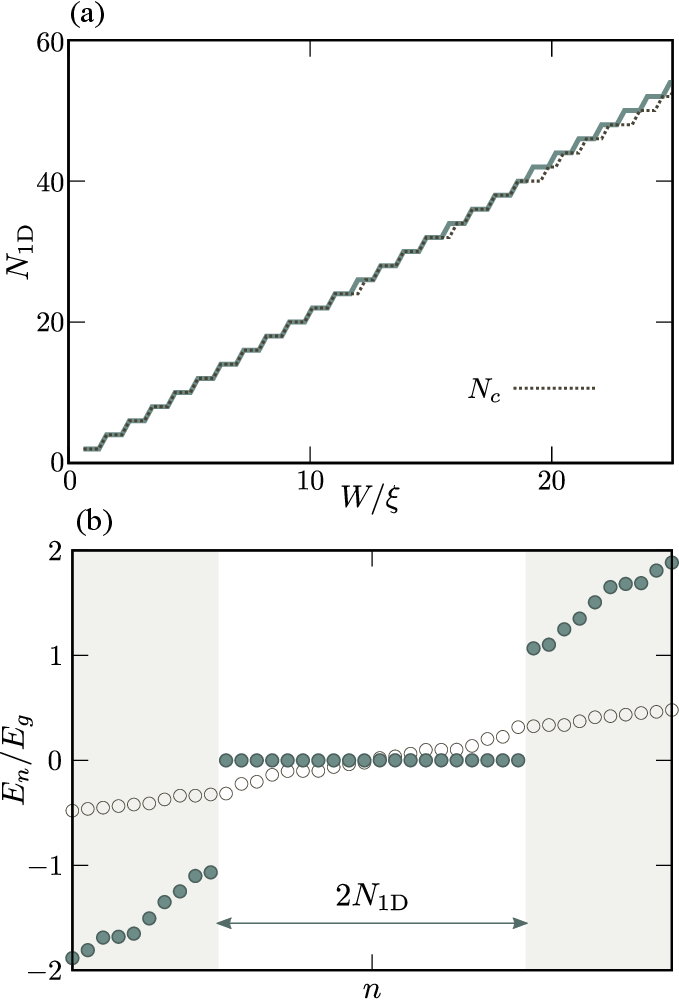}
\caption{(a) 1D winding number, $N_{\text{1D}}$, as a function $W$, with dotted line representing the number of the propagating channels $N_c$.
(b) Energy spectrum under open boundary conditions in both $x$ and $y$ directions, where the colored (white) dots indicate the result for $\lambda=0.5t$ ($\lambda=0$).
Each dot corresponds to two-fold degenerate states due to Kramars degeneracy.
For $\lambda=0.5t$, we clearly observe the emergence of $N_{\text{1D}}$-fold degenerate ZESs.
}
\label{fig:figure2}
\end{center}
\end{figure}

The main purpose of this Letter is to uncover finite-size-induced topological phenomena of superconducting systems.
We consider a thin film exhibiting a persistent spin texture (PST)~\cite{bernevig_06}, fabricated on a high-$T_c$ cuprate $d_{xy}$-wave SC [see Fig.~\ref{fig:figure1}(a)].
Specifically, we assume that a unidirectional spin-orbit coupling (SOC), which induces the PST, and a proximity-induced $d_{xy}$-wave pair potential coexist in the thin-film.
A 2D $d_{xy}$-wave SC is a prototypical nodal topological SC, hosting flat-band ABSs at its edges~\cite{bruder_90,hu_94,tanaka_95,asano_04,sato_11} [see Fig.~\ref{fig:figure1}(b)].
Reducing the system size in one direction induces finite-size effects, resulting in a characteristic energy dispersion in the ABS spectrum,
where multiple band crossings occur at zero energy [see Fig.~\ref{fig:figure1}(c)].
Furthermore, introducing the unidirectional SOC of the PST opens a full energy gap in the ABS spectrum [see Fig.~\ref{fig:figure1}(d)].
The resulting gapped quasi-one-dimensional (Q1D) system falls into the chiral symmetry class (class AIII) of the Altland-Zirnbauer (AZ) classification~\cite{schnyder_08}.
This class allows a $\mathbb{Z}$ topological invariant in one dimension, namely a one-dimensional (1D) winding number.
This winding number can take unusually large values, which almost coincides with the number of the propagating channels in the Q1D system [see Fig.~\ref{fig:figure2}(a)].
The appearance of highly degenerate zero-energy edge states (ZESs), associated with this exceptionally large winding number, is confirmed [see Fig.~\ref{fig:figure2}(b)].
We demonstrate that these ZESs lead to prominent zero-bias conductance quantization in \emph{dirty} normal-metal--SC (NM--SC) junctions (see Fig.~\ref{fig:figure5}).
This disorder-independent perfect charge transport is analogous to the anomalous proximity effects of 2D spin-triplet $p_x$-wave SCs that host flat-band Majorana bound states 
~\cite{tanaka_04,tanaka_05(1),asano_07,ikegaya_15,ikegaya_16(1),ikegaya_21,lee_25,ikegaya_25,asano_06(1),asano_06(2),ikegaya_16(2)}.

High-$T_c$ cuprates are employed as the parent SC due to their large superconducting gap~\cite{shen_03,shen_17},
which implies that the induced energy gap in ABSs will be sufficiently large for experimental observation.
The PST was originally discovered in zinc-blende semiconductor quantum wells~\cite{schliemann_17,kohda_17}.
More recently, the formation of the PST has been predicted in a variety of 2D materials,
including wurtzite-structured semiconductors~\cite{saito_15}, ferroelectric semiconductors~\cite{tsymbal_18,ghosez_19,picozzi_19,jin_20,ren_20,rondinelli_20,hao_22},
monolayer transition metal dichalcogenides~\cite{wang_19,ishii_20}, and group IV-V compounds~\cite{santoso_22},
enabling the experimental realization of the proposed system.
Consequently, we propose an intriguing route for exploring novel topological superconducting phases driven by gapped ABSs.

\textit{Emergence of zero-energy states.}
We describe the present system using a 2D tight-binding model on a square lattice.
A lattice site is indicated by a vector $\boldsymbol{r}=j \boldsymbol{x} + m \boldsymbol{y}$, where $|\boldsymbol{x}|=|\boldsymbol{y}|=a$.
We apply a periodic boundary condition in the $x$ direction, while an open boundary condition is applied in the $y$ direction;
the thin film is placed on $1 \leq m \leq M$ with $W=Ma$ representing the width of the system.
The Bogoliubov--de Gennes (BdG) Hamiltonian reads,
\begin{align}
\begin{split}
&H=\frac{1}{2}\sum_{k_x} C^{\dagger}_{k_x} H_{k_x} C_{k_x},\\
&C^{\dagger}_{k_x}=[c^{\dagger}_{k_x,\uparrow},c^{\dagger}_{k_x,\downarrow},c^{\mathrm{T}}_{-k_x,\uparrow},c^{\mathrm{T}}_{-k_x,\downarrow}],\\
&c_{k_x,s}=[c_{k_x,1,s},c_{k_x,2,s},\cdots,c_{k_x,M,s}]^{\mathrm{T}},\\
&H_{k_x} = \xi_{k_x}\otimes \tau_z + \Lambda \otimes s_z - \Delta_{k_x} \otimes s_y \otimes \tau_y,\\
&\xi_{k_x}=-2t\cos(k_x a)-\mu + 4t -t A_+,\\
&\Lambda = \frac{i \lambda}{2} A_-,\quad
\Delta_{k_x}=\frac{i \Delta}{2}\sin (k_x a) A_-,
\label{eq:bdgham}
\end{split}
\end{align}
where $A_{\pm}$ is the $(M\times M)$ matrices with
\begin{align}
(A_{\pm})_{ij}=\left\{ \begin{array}{cl}
1 & \text{for } i=j+1\\ \pm 1 & \text{for } i=j-1 \\ 0 & \text{otherwise}\end{array} \right.,
\end{align}
$c_{k_x,m,s}$ is an annihilation operator of an electron at $y=ma$ with momentum $k_x$ and spin $s$ ($=\uparrow,\downarrow$),
$t$ denotes the nearest-neighbor hopping integral,
$\mu$ is the chemical potential,
$\lambda$ represents the strength of the unidirectional SOC forming the PST,
and $\Delta$ denotes the proximity-induced spin-singlet $d_{xy}$-wave pair potential.
$\tau_{\nu}$ and $s_{\nu}$ for $\nu=x,y,z$ denote the Pauli matrices in Nambu and spin spaces, respectively.
When we apply a periodic boundary condition in both $x$ and $y$ directions, the BdG Hamiltonian is rewritten as,
$H_{\boldsymbol{k}} = \xi_{\boldsymbol{k}}\tau_z + \lambda \sin(k_y a) s_z-\Delta_{\boldsymbol{k}} s_y \otimes \tau_y$
with $\xi_{\boldsymbol{k}}=-2t\cos(k_x a)-2t \cos(k_y a) -\mu+4t$ and $\Delta_{\boldsymbol{k}} = \Delta \sin(k_x a) \sin(k_y a)$.
The PST inherently guarantees spin-rotational symmetry~\cite{schliemann_17,kohda_17}:
\begin{align}
\left[ R_z, H_{k_x}\right] =0, \quad R_z = is_z \otimes   \tau_z,
\end{align}
where $R_z$ describes the spin rotation along the $z$ axis.
Therefore, we can decompose the Hamiltonian into two subsectors, each corresponding to a different eigenvalue of $R_z$:
\begin{align}
\begin{split}
&H=\frac{1}{2}\sum_{k_x,s} C^{\dagger}_{k_x,s} H_{k_x,s} C_{k_x,s},\\
&C^{\dagger}_{k_x,s}=[c^{\dagger}_{k_x,s},c^{\mathrm{T}}_{-k_x,\bar{s}}],\\
&H_{k_x,s}=\left[\begin{array}{cc} \xi_{k_x} + \sigma_s \Lambda & \sigma_s \Delta_{k_x} \\ \sigma_s \Delta_{k_x} & -\xi_{k_x} - \sigma_s \Lambda \end{array}\right],
\end{split}
\end{align}
where $\bar{s}$ represents the opposite spin of $s$, and $\sigma_{s} = +1 (-1)$ for $s=\uparrow (\downarrow)$.
In the \emph{absence} of the SOC (i.e., $\lambda=0$), the Hamiltonian $H_{k_x,s}$ belongs to class CI of the AZ classification:
\begin{align}
\begin{split}
&T H_{k_x,s} T^{-1} = H_{-k_x,s},\quad  T=\mathcal{K},\\
&C H_{k_x,s} C^{-1} = -H_{-k_x,s},\quad  C=\tau_y \mathcal{K},\\
&\Gamma H_{k_x,s} \Gamma^{-1} = -H_{k_x,s},\quad  \Gamma=\tau_y,
\end{split}
\end{align}
where $\mathcal{K}$ denotes the complex conjugation operator, and the symmetry operators satisfy $T^2=1$, $C^2=-1$, and $\Gamma^2=1$.
However, in the \emph{presence} of the SOC (i.e., $\lambda\neq0$), the Hamiltonian falls into class AIII, preserving only the chiral symmetry $\Gamma$.
According to the AZ classification~\cite{schnyder_08}, a \emph{fully gapped} energy spectrum of $H_{k_x,s}$ allows us to define the 1D winding number:
\begin{align}
N_{\text{1D}}= \frac{i}{4 \pi} \sum_{s=\uparrow,\downarrow} \int dk_x \mathrm{Tr}[\Gamma H_{k_x,s}^{-1} \partial_{k_x} H_{k_x,s} ] \in \mathbb{Z},
\label{eq:wind}
\end{align}
where the trace is taken over both the Nambu space and the lattice sites along the $y$ direction.
The unidirectional SOC is essential for defining the 1D winding number $N_{\text{1D}}$, as generic SOCs such as Rashba SOC break spin-rotation symmetry.
In this case, the Q1D systems are in class DIII, and the AZ classification changes from $\mathbb{Z}$ to $\mathbb{Z}_2$.

Figures~\ref{fig:figure1}(b)--\ref{fig:figure1}(d) present the energy eigenvalues of $H_{k_x,s}$ as a function of $k_x$,
which are obtained by diagonalizing $H_{k_x,s}$ numerically.
The parameters are set as $\mu=t$ and $\Delta=0.2t$.
The superconducting coherence length is evaluated as $\xi=1/\{ \pi \mathrm{artanh}(\Delta/2t) \}\sim3.17a$.
Note that the energy eigenvalues of $H_{k_x,\uparrow}$ and those of $H_{k_x,\downarrow}$ are completely overlapped due to Kramers degeneracy.
In Fig.~\ref{fig:figure1}(b), we plot the energy eigenvalues with $\lambda=0$, where the system width is sufficiently larger than the superconducting coherence length, $W=800a\sim252\xi$.
In this regime, the flat-band ABSs are crearly visible at zero energy.
In Fig.~\ref{fig:figure1}(c), we show the energy eigenvalues with $\lambda=0$ in a narrower system with $W=30a\sim9.46\xi$.
The finite-size effects leads to hybridization between the ABSs localized at $y/a=1$ and $y/a=W/a$, resulting in the breakdown of the flat-band structure.
Nevertheless, multiple band crossings persist at zero energy.
As detailed in the Supplemental Material (SM)~\cite{sm}, these band crossings are protected by inversion symmetry~\cite{kobayashi_14,timm_17},
which holds only in the absence of the SOC (i.e., $\lambda=0$).
Consequently, introducing the unidirectional SOC is expected to open a full energy gap in the ABS spectrum.
This is indeed observed in Fig.~\ref{fig:figure1}(d), where the SOC is introduced with $\lambda=0.5t$.
Since $H_{k_x,s}$ now describes the gapped systems in class AIII, the 1D winding number $N_{\text{1D}}$ in Eq.~(\ref{eq:wind}) becomes well defined.

In Fig.~\ref{fig:figure2}(a), the 1D winding number $N_{\rm 1D}$ is plotted as a function of the system width $W$, with $\lambda=0.5t$.
The results show that $N_{\rm 1D}$ increases with $W$ and reaches values substantially larger than unity.
For instance, at $W=30a\sim9.46\xi$, the winding number reaches $N_{\rm 1D}=20$.
The dotted line denotes the number of the propagating channels $N_c=2n_c$,
where $n_c$ is the largest integer satisfying $\mu > 2t\{1-\cos(\frac{n_c\pi}{W+1})\}$.
The factor of two reflects spin degeneracy.
Notably, $N_{\rm 1D}$ closely tracks $N_c$, and the underlying mechanism is discussed later.
The bulk-boundary correspondence guarantees the presence of $N_{\rm 1D}$-fold degenerate ZESs at the system edge.
In Fig.~\ref{fig:figure2}(b), we show the energy eigenvalues under open boundary conditions in both $x$ and $y$ directions, where the system is placed on $1 \leq j \leq L/a$.
We choose $W=30a\sim9.46\xi$ and $L=600a\sim189.27\xi$.
The colored (white) dots indicate the result for $\lambda=0.5t$ ($\lambda=0$); each dot represents a two-fold degenerate state caused by Kramers degeneracy.
In the absence of SOC, the energy spectrum remains gapless.
 When SOC is present, $2N_{\rm 1D}$ ZESs appear inside the finite energy gap $E_g$, where the factor of two arises from both edges in the $x$ direction.
Under the present parameters, the system hosts 40 ZESs.
The winding number is limited by the number of Fermi points in the 1D Brillouin zone~\cite{sato_11}, which typically results in small values. 
In contrast, the proposed Q1D system exhibits multiple band crossings due to the finite-size effect and consequently supports an exceptionally large number of the topologically protected ZESs.

We examine the size of the induced energy gap, $E_g$, as also indicated in Fig.~\ref{fig:figure1}(d).
In Fig.~\ref{fig:figure2}(a), $E_g$ is plotted as a function of the system width $W$.
The solid and dashed lines correspond to $\lambda=0.5t$ and $\lambda=t$, respectively.
As $W$ increases, the induced gap $E_g$ exhibits oscillatory decay, with larger values of $\lambda$ yielding larger gaps.
We now adopt Bi$_2$Sr$_2$CaCu$_2$O$_8$ as the parent SC, which has a large superconducting gap, $\Delta \sim$ 30~meV~\cite{shen_03,shen_17}.
With its in-plane lattice constant $a \sim 5.4$~\AA, the superconducting coherent length is estimated as $\xi \sim 1.71$~nm.
As an example, when $W=64a \sim 20.21\xi \sim 34.6$~nm and $\lambda a=t a=5\Delta a \sim 810$~meV\AA, the resulting induced gap is $E_g=0.044\Delta \sim 1.3$~meV,
which remains sufficiently large for experimental observation.
Moreover, compatible or even stronger SOCs, reaching several eV\AA, have been theoretically predicted in various 2D systems with a PST
(see, for example, Refs.~\cite{ishii_20,santoso_22}).
Thus, while a more microscopic evaluation of $E_g$ remains an important future task,
these results suggest that topological phases protected by the induced gap of ABSs could be experimentally accessible.
\begin{figure}[t]
\begin{center}
\includegraphics[width=0.375\textwidth]{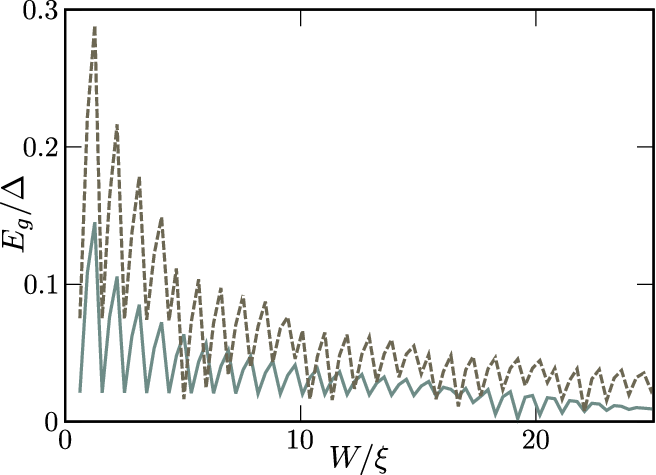}
\caption{Induced energy gap, $E_g$, as a function of the system width, $W$, where the solid (dashed) line shows the result for $\lambda=0.5t$ ($\lambda=t$).
}
\label{fig:figure3}
\end{center}
\end{figure}

\textit{Effective theory.}
To obtain an intuitive understanding of the numerical results, we derive a low-energy effective Hamiltonian.
For this purpose, we consider the BdG equation in the continuum limit:
\begin{align}
H_{s}(k_x,y)\psi_{s}(k_x,y) = E \psi_{s}(k_x,y),
\end{align}
with
\begin{align}
\begin{split}
&H_{s}(k_x,y)=H^{d_{xy}}_{s}(k_x,y) + H^{\lambda}_{s}(y),\\
&H^{d_{xy}}_{s}(k_x,y)=\left[\begin{array}{cc} \xi(k_x,y) & \sigma_s \Delta(k_x,y) \\ \sigma_s \Delta(k_x,y) & -\xi(k_x,y) \end{array}\right],\\
&H^{\lambda}_{s}(y)=\sigma_s \left[\begin{array}{cc} \Lambda(y) & 0 \\ 0 & - \Lambda(y) \end{array}\right],\\
&\xi(k_x,y) = -\frac{\hbar^2}{2m}\partial_y^2 - \tilde{\mu}, \quad \tilde{\mu}=\mu-\frac{\hbar^2k_x^2}{2m},\\
&\Delta(k_x,y) = -i \frac{\tilde{\Delta}}{k_F} \partial_y, \quad \tilde{\Delta}=\frac{\Delta k_x}{k_F}, \quad
\Lambda(y) = -i\tilde{\lambda} \partial_y,
\end{split}
\end{align}
where $ta^2=\hbar^2/(2m)$, $\lambda a=\tilde{\lambda}$, and $k_F=\sqrt{2m\mu}/\hbar$ denotes the Fermi wave number.
We first explore the wave functions of the ABSs in the absence of the SOC, satisfying:
\begin{align}
H^{d_{xy}}_{s}(k_x,y) \psi^{\pm}_{s}(k_x,y) = \pm \epsilon_{s}(k_x)\psi^{\pm}_{s}(k_x,y),
\end{align}
where $\pm \epsilon_{s}(k_x)$ represents the energy eigenvalues of the ABSs, as also illustrated in Fig.~\ref{fig:figure3}(a).
While obtaining $\epsilon_{s}(k_x)$ and $\psi^{\pm}_{s}(k_x,y)$ across the entire range of $k_x$ is challenging, as detailed in the SM~\cite{sm},
we can analytically solve the BdG equation at momenta where zero-energy band crossings occur: $H^{d_{xy}}_{s}(k_n,y) \psi^{\pm}_{s}(k_n,y) = 0$.
Specifically, the zero-energy band crossings occur at momenta:
\begin{align}
\begin{split}
&k_n=\mathrm{sgn}[n] \sqrt{\frac{k_F^2-q_n^2}{1+\left(\frac{\Delta}{2\mu}\right)^2}},\\
&n=\pm1,\pm2,\cdots,\pm n_c,
\end{split}
\end{align}
with the corresponding zero-energy wave functions:
\begin{align}
\begin{split}
&\psi^{\pm}_{s}(k_n,y)=\left\{ \begin{array}{cl}
e^{\mp i\frac{\pi}{4}}\varphi^{\pm}_{s}(k_n,y) & \text{for } |n| \in \text{odd} \\ e^{\mp i\frac{\pi}{4}} \varphi^{\mp}_{s}(k_n,y) & \text{for } |n| \in \text{even}\end{array} \right.,\\
&\varphi^{\pm}_{s}(k_n,y)=\frac{1}{2}\left[ \begin{array}{cc}
\phi_{s,+}(k_n,y) \pm \phi_{s,-}(k_n,y) \\ i \left\{ \phi_{s,+}(k_n,y) \mp \phi_{s,-}(k_n,y) \right\} \end{array}\right],\\
&\phi_{s,\pm}(k_n,y)=c_{k_n,s,\pm}\sin (q_n y) e^{\pm s \kappa_n y},\\
&q_n=\frac{n\pi}{W},\quad \kappa_n = \frac{\Delta}{2\mu}k_n,\quad
c_{s,\pm,n}=\sqrt{\frac{2X_n}{W}e^{\mp s \kappa_n W}}\\
&X_n=\frac{\kappa_nW}{\sinh (\kappa_n W)} \left\{ 1+\left(\frac{\Delta}{2\mu}\right)^2\frac{k_n^2}{q_n^2}\right\},
\label{eq:zero_wave}
\end{split}
\end{align}
where $n_c$ in the continuum limit is given by $n_c=[Wk_F/\pi]_G$ with $[\cdots]_G$ representing the Gauss symbol, which takes the integer part of the argument.
A detailed derivation for Eq~(\ref{eq:zero_wave}) is provided in the SM.
Notably, the total number of zero-energy band crossings coincides with the number of propagating channels, $N_c=2n_c$ [see Fig.~\ref{fig:figure3}(a)].
To proceed with our analysis, we treat the SOC, $H^{\lambda}_{s}(y)$, as a perturbation and construct the low-energy effective Hamiltonian as follows:
\begin{align}
\begin{split}
&H^{\mathrm{eff}}_s(k_x)=\left[\begin{array}{cc} A_{++} & A_{+-} \\ A_{-+} & A_{--} \end{array}\right] + O(\lambda^2),\\
&A_{\eta\eta^{\prime}}=\int dy \left\{ \psi^{\eta}_{s}(k_x,y) \right\}^{\dagger}H_{s}(k_x,y) \psi^{\eta^{\prime}}_{s}(k_x,y),
\end{split}
\end{align}
where $A_{\eta\eta}=\eta \epsilon_{s}(k_x)$.
In the vicinity of $k_x=k_n$, we obtain, $H^{\mathrm{eff}}_s(k_n+\delta k)=H^{\mathrm{eff}}_{n,s}(\delta k) + O(\delta k^2)$, with
\begin{align}
\begin{split}
&H^{\mathrm{eff}}_{n,s}(\delta k)=\left\{ \begin{array}{cl}
\left[\begin{array}{cc} v_n \delta k & m_n \\ m_n & -v_n \delta k \end{array}\right] & \text{for } |n| \in \text{odd} \\
&\\
\left[\begin{array}{cc} -v_n \delta k & -m_n \\ -m_n & v_n \delta k \end{array}\right] & \text{for } |n| \in \text{even}\end{array} \right.,\\
& v_n=\frac{\hbar^2k_n}{m}\left\{1+\left(\frac{\Delta}{2\mu}\right)^2\right\}X_n,\quad
m_n= \tilde{\lambda} \kappa_n X_n,
\label{eq:eff_ham}
\end{split}
\end{align}
where $m_n$ represents the mass term that opens the energy gap in the ABS spectrum, as depicted in Fig.~\ref{fig:figure3}(b).
A detailed derivation for Eq.~(\ref{eq:eff_ham}) is provided in the SM.
The low-energy effective Hamiltonian preserves chiral symmetry:
\begin{align}
\gamma H^{\mathrm{eff}}_s(k_x) \gamma^{-1}=-H^{\mathrm{eff}}_s(k_x), \quad \gamma=\left[\begin{array}{cc}0&-i\\i&0 \end{array}\right],
\end{align}
which allows us to define the winding number:
\begin{align}
N_{\text{1D}}^{\mathrm{eff}} = \frac{i}{4 \pi} \sum_{s} \int dk_x \mathrm{Tr}[\gamma \{ H^{\mathrm{eff}}_s(k_x)\}^{-1} \partial_{k_x}  H^{\mathrm{eff}}_s(k_x) ].
\end{align}
On the basis of Ref.~\cite{sato_11}, we can compute $N_{\text{1D}}^{\mathrm{eff}}$ using only the Hamiltonian in the vicinity of $k_x=k_n$:
\begin{align}
N_{\text{1D}}^{\mathrm{eff}} = \frac{1}{2} \sum_s  \sum_{n=-n_c}^{n_c} \mathrm{sgn}[v_n] \mathrm{sgn}[m_n],
\label{eq:wind2}
\end{align}
where the detailed derivation of Eq.~(\ref{eq:wind2}) is provided in the SM.
Applying this formula, we eventually obtain
\begin{align}
N_{\text{1D}}^{\mathrm{eff}} =N_c,
\end{align}
which shows excellent agreement with the numerical result shown in Fig.~\ref{fig:figure2}(a).
Consequently, our effective theory confirms that the nontrivial topology of the present systems indeed originates from the energy gap of the ABSs.
\begin{figure}[t]
\begin{center}
\includegraphics[width=0.375\textwidth]{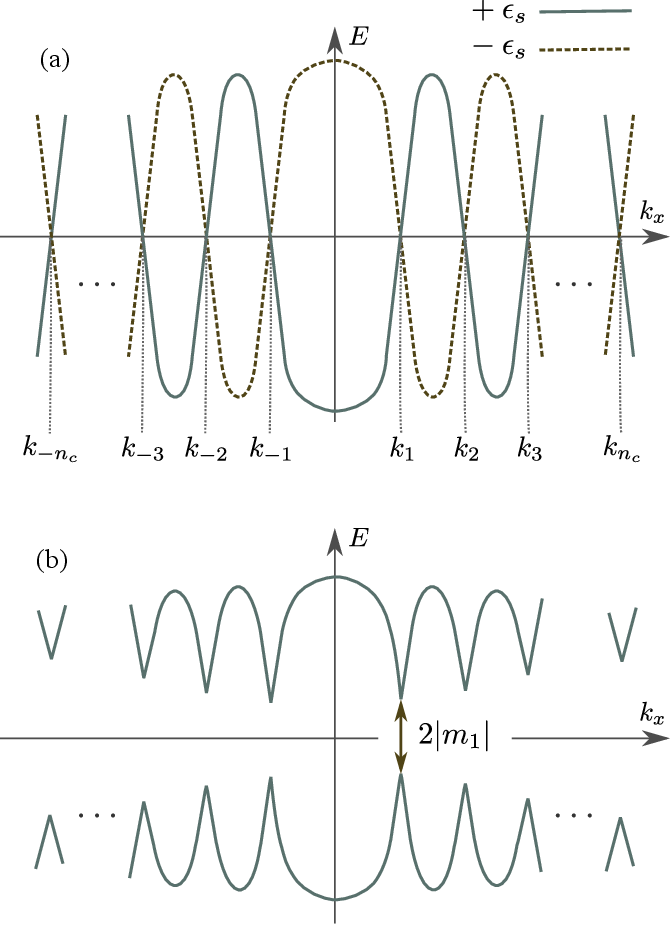}
\caption{Schematic of the energy spectra of the ABSs for (a) $\tilde{\lambda}=0$ and for (b) $\tilde{\lambda} \neq 0$, described by the effective Hamiltonian in Eq.~(\ref{eq:eff_ham}).}
\label{fig:figure4}
\end{center}
\end{figure}

\textit{Anomalous proximity effect.}
\begin{figure}[t]
\begin{center}
\includegraphics[width=0.5\textwidth]{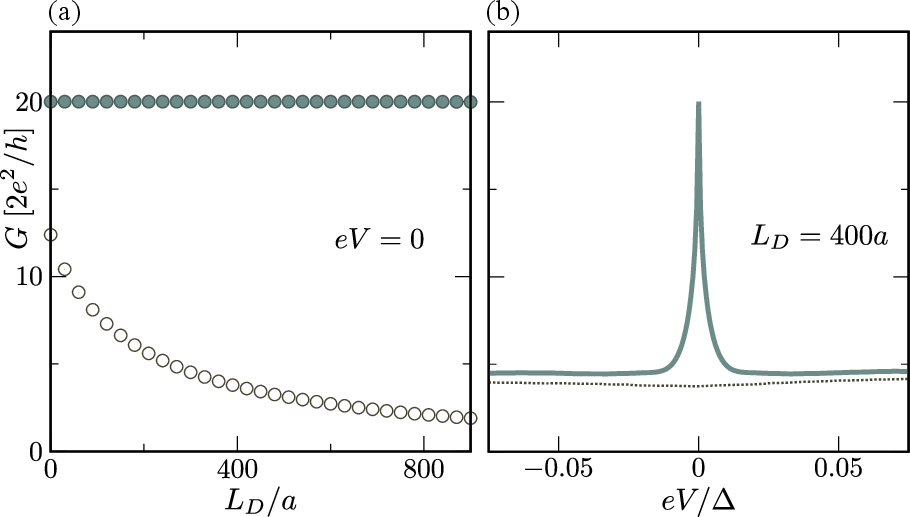}
\caption{(a) Zero-bias conductance as a function of the length of the dirty NM segment, $L_D$, with the colored (white) dots representing the result for $\lambda=0.5t$ ($\lambda=0$).
(b) Conductance spectra at $L_D=400a$, with the solid (dashed) line representing the result for $\lambda=0.5t$ ($\lambda=0$).
For the finite-size-induced topological phase, (i.e., $\lambda=0.5t$), we observe zero-bias conductance quantization at $(2e^2/h) N_{\text{1D}}$, independent of $L_D$.}
\label{fig:figure5}
\end{center}
\end{figure}
Finally, we describe anomalous charge transport induced by the $N_{\text{1D}}$-fold degenerate ZESs.
We consider a junction consisting of three segments:
a ballistic NM segment for $-\infty \leq j < 1$, a dirty NM segment for $1 \leq j \leq L_D/a$, and a SC segment for $L_D/a< j \leq \infty$.
For numerical calculations, the BdG Hamiltonian in Eq.~(\ref{eq:bdgham}) is reformulated in a real-space basis by performing a Fourier transformation along the $x$ direction:
$c_{k_x,m,s} \rightarrow c_{\boldsymbol{r},s}$.
 The proximity-induced pair potential is assumed to exist only in the SC segment, i.e., $j>L_D/a$.
For the dirty NM segment, we introduce a non-magnetic random on-site potential:
\begin{align}
H_D = \sum_{j=1}^{L_D/a}\sum_{m,s}v_{\boldsymbol{r}} c^{\dagger}_{\boldsymbol{r},s}c_{\boldsymbol{r},s},
\end{align}
where $v_{\boldsymbol{r}}$ is given randomly in the range $-v_{\mathrm{imp}}/2 \leq v_{\boldsymbol{r}} \leq v_{\mathrm{imp}}/2$.
We calculate the differential conductance at zero temperature on the basis of the Blonder--Tinkham--Klapwijk formalism~\cite{klapwijk_82}:
\begin{align}
G(eV) = \frac{e^{2}}{h} \sum_{\zeta,\eta}
\left[ \delta_{\zeta,\eta} - \left| r^{ee}_{\zeta,\eta} \right|^{2}
+ \left| r^{he}_{\zeta,\eta} \right|^{2} \right]_{E=eV},
\end{align}
where $r^{ee}_{\zeta,\eta}$ and $r^{he}_{\zeta,\eta}$ denote a normal and an Andreev reflection coefficient at energy $E$, respectively.
The indices $\zeta$ and $\eta$ correspond to the outgoing and incoming channels.
These reflection coefficients are calculated using recursive Green's function techniques~\cite{fisher_81,ando_91}.
For numerical calculations, the parameters are fixed as $\mu=t$, $\Delta=0.2t$, $v_{\mathrm{imp}}=0.5t$, and  $W=30a$.
The conductance is averaged over $5\times10^3$ samples with different random potential configurations.

In Fig.~\ref{fig:figure5}(a), we plot the differential conductance at zero bias as a function of the length of the dirty NM segment.
The colored (white) dots correspond to $\lambda=0.5t$ ($\lambda=0$).
In the absence of SOC, the zero-bias conductance decreases monotonically with increasing $L_D$.
In contrast, for $\lambda=0.5t$, the zero-bias conductance remains perfectly quantized, independent of $L_D$:
\begin{align}
G(0) = \frac{2e^2}{h} \times N_{\text{1D}} =\frac{2e^2}{h} \times N_c,
\end{align}
where $N_{\text{1D}}=20$ for the given parameters.
Moreover, the conductance spectrum for $\lambda=0.5t$ exhibits a prominent zero-bias peak, as shown in Fig.~\ref{fig:figure5}(b).
This perfect conductance quantization suggests that the $N_{\text{1D}}$-fold degenerate ZESs, initially localized at the junction interface,
penetrate into the dirty NM segment while retaining their high degree of degeneracy.
Therefore, they form perfect transmission channels within the dirty NM-SC junction~\cite{ikegaya_15,ikegaya_16(1)}.
This unusual superconducting proximity effect, accompanied by the penetration of topologically protected ZESs, is analogous to the anomalous proximity effect observed in 2D nodal $p_x$-wave SCs%
~\cite{tanaka_04,tanaka_05(1)}.
As a result, we propose a striking perfect charge transport that clearly manifests the presence of highly degenerate ZESs.

\textit{Summary.}
In summary, we investigate a novel topological superconducting phase emerging in a Q1D system, where a unidirectional SOC potential of a PST coexists with a $d_{xy}$-wave pair potential.
Finite-size effects lead to hybridization of the flat-band ABSs, resulting in multiple band crossings.
Introducing the unidirectional SOC opens a full energy gap in the ABS spectrum and gives rise to an exceptionally large winding number, $N_{\text{1D}}$.
Numerical calculations demonstrate the emergence of topologically protected $N_{\text{1D}}$-fold degenerate ZESs at both ends of the Q1D system.
Furthermore, a distinctive perfect transport phenomenon serves as an unambiguous signature of these finite-size-induced topological phases.
These findings offer new avenues for studying topological superconducting phases driven by gapped ABSs.

\begin{acknowledgments}
S.I. is supported by a Grant-in-Aid for Early-Career Scientists (JSPS KAKENHI Grant No. JP24K17010). 
S. K. was supported by JSPS KAKENHI (Grants No. JP22K03478 and No. JP24K00557) and JST CREST (Grant No. JPMJCR19T2).
\end{acknowledgments}

\section*{Data availability}
The data that support the findings of this article are openly available \cite{data}.

\pagebreak
\onecolumngrid
\begin{center}
  \textbf{\large Supplemental Material for \\ ``Exceptionally large winding number of a finite-size topological superconductor''}\\ \vspace{0.3cm}
Satoshi Ikegaya$^{1,2}$, Shingo Kobayashi$^{3}$, and Yasuhiro Asano$^{4}$ \\ \vspace{0.1cm}
{\itshape 
$^{1}$Institute for Advanced Research, Nagoya University, Nagoya 464-8601, Japan\\
$^{2}$Department of Applied Physics, Nagoya University, Nagoya 464-8603, Japan\\
$^{3}$RIKEN Center for Emergent Matter Science, Wako, Saitama, 351-0198, Japan\\
$^{4}$Department of Applied Physics, Hokkaido University, Sapporo 060-8628, Japan}
\date{\today}
\end{center}

\section{Zero-energy Band crossings protected by inversion symmetry}
In this section, we discuss the relationship between the zero-energy band crossings in Fig.~1(c) and inversion symmetry of the Hamiltonian.
We consider the Bogoliubov--de Gennes (BdG) Hamiltonian in Eq.~(4) of the main text:
\begin{align}
\begin{split}
&H_{k_x,s}=\left[\begin{array}{cc} \xi_{k_x} + \sigma_s \Lambda & \sigma_s \Delta_{k_x} \\ \sigma_s \Delta_{k_x} & -\xi_{k_x} - \sigma_s \Lambda \end{array}\right],\\
&\xi_{k_x}=-2t\cos(k_x a)-\mu + 4t -t A_+,\\
&\Lambda = \frac{i \lambda}{2} A_-,\quad
\Delta_{k_x}=\frac{i \Delta}{2}\sin (k_x a) A_-,
\end{split}
\end{align}
where $A_{\pm}$ is the $(M\times M)$ matrices with
\begin{align}
(A_{\pm})_{ij}=\left\{ \begin{array}{cl}
1 & \text{for } i=j+1\\ \pm 1 & \text{for } i=j-1 \\ 0 & \text{otherwise}\end{array} \right..
\end{align}
In the \emph{absence} of the spin-orbit coupling (SOC) potential (i.e., $\lambda=0$), the Hamiltonian $H_{k_x,s}$ preserves inversion symmetry:
\begin{align}
\begin{split}
&PH_{k_x,s}P^{-1}=H_{-k_x,s},\quad
P= \left[\begin{array}{cc} P_y & 0 \\ 0 & P_y \end{array}\right],\\
&(P_y)_{ij}=\left\{ \begin{array}{cl}
1 & \text{for } i=M-j+1 \\ 0 & \text{otherwise}\end{array} \right..
\label{eq:inv-symm}
\end{split}
\end{align}
This Hamiltonian also preserves particle-hole-like symmetry:
\begin{align}
C_P H_{k_x,s} C_P^{-1} = -H_{-k_x,s},\quad  C_P= \Xi \mathcal{K}, \quad
\Xi= \left[\begin{array}{cc} 0 & P_y \\ P_y & 0 \end{array}\right],
\end{align}
where $\mathcal{K}$ denotes the complex conjugation operator, and the operator $C_P$ satisfies $C_P^2 = +1$.
Combining these symmetries, we find $CP$-like symmetry as:
\begin{align}
U_{CP} H^{\mathrm{T}}_{k_x,s} U_{CP}^{-1} = -H_{k_x,s},\quad  U_{CP}= \Xi P^{\ast} = \tau_x,
\label{eq:cp-symm}
\end{align}
where $(C_P P)^2=U_{CP} U_{CP}^{\ast}=+1$ and hence $U_{CP}^{\mathrm{T}}=U_{CP}$.
Following Refs.~\cite{kobayashi_14,timm_17}, we define a $\mathbb{Z}_2$ invariant that characterizes the zero-energy band crossings in the spectrum of  $H_{k_x,s}$ at $\lambda=0$.
Since $U_{CP}$ is a symmetric matrix, it can be decomposed as $U_{CP}=V Q V^{\mathrm{T}}$, where $Q$ is a diagonal matrix and $V$ is a unitary matrix.
We define $\Omega = \sqrt{Q}^{\dagger} V^{\dagger}$, where $\sqrt{Q}$ is well-defined due to the diagonal nature of $Q$.
Using $\Omega$, we rewrite Eq.~(\ref{eq:cp-symm}) as:
\begin{align}
\Omega^{\ast} H^{\mathrm{T}}_{k_x,s} \Omega^{\mathrm{T}} = - \Omega H _{k_x,s} \Omega^{\dagger}.
\end{align}
As a result, we find that the transformed Hamiltonian, $\tilde{H} _{k_x,s}=\Omega H _{k_x,s} \Omega^{\dagger}$, becomes an antisymmetric matrix:
\begin{align}
\tilde{H} _{k_x,s}^{\mathrm{T}}=-\tilde{H} _{k_x,s}.
\end{align}
The antisymmetric nature of $\tilde{H} _{k_x,s}$ allows us to define a $\mathbb{Z}_2$ invariant using its Pfaffian:
\begin{align}
(-1)^{\nu} = \mathrm{sgn}\left[ \mathrm{Pf}(\tilde{H} _{k',s}) \mathrm{Pf}(\tilde{H} _{k'',s}) \right]
\end{align}
where $k'$ and $k''$ are two arbitrary momenta.
When $\nu=1$ mod $2$, there is an odd number of zero-energy band crossings between $k'$ and $k''$~\cite{kobayashi_14,timm_17}.
Thus, the presence of zero-energy band crossings in the present system is protected by a nontrivial $\mathbb{Z}_2$ invariant:
\begin{align}
\mathrm{sgn}\left[ \mathrm{Pf}(\tilde{H} _{k_n+\delta k,s}) \mathrm{Pf}(\tilde{H} _{k_n-\delta k,s}) \right] = -1
\end{align}
where $k_n$ denotes the momenta at which zero-energy band crossing occurs and $\delta k \ll 1$.
These band crossings can only be removed by perturbations that break the $CP$-like symmetry in Eq.~(\ref{eq:cp-symm}).
In our model, such symmetry breaking is achieved by introducing the SOC term (i.e., $\lambda \neq 0$), which inherently breaks the inversion symmetry in Eq.~(\ref{eq:inv-symm}).
\vspace{22pt}

\section{Effective Theory}
\subsection{Low-energy effective Hamiltonian}
We derive an effective Hamiltonian for a quasi-one dimensional system with a unidirectional SOC potential of a persistent spin texture and a $d_{xy}$-wave pair potential.
We begin with the BdG equation in Eq.~(7) of the main text:
\begin{align}
\begin{split}
&H_{s}(k_x,y)\psi_{s}(k_x,y) = E \psi_{s}(k_x,y), \\
&H_{s}(k_x,y)=H^{d_{xy}}_{s}(k_x,y) + H^{\lambda}_{s}(y),\\
&H^{d_{xy}}_{s}(k_x,y)=\left[\begin{array}{cc} \xi(k_x,y) & \sigma_s \Delta(k_x,y) \\ \sigma_s \Delta(k_x,y) & -\xi(k_x,y) \end{array}\right],\\
&H^{\lambda}_{s}(y)=\sigma_s \left[\begin{array}{cc} \Lambda(y) & 0 \\ 0 & - \Lambda(y) \end{array}\right],\\
&\xi(k_x,y) = -\frac{\hbar^2}{2m}\partial_y^2 - \tilde{\mu}, \quad \tilde{\mu}=\mu-\frac{\hbar^2k_x^2}{2m},\\
&\Delta(k_x,y) = -i \frac{\tilde{\Delta}}{k_F} \partial_y, \quad \tilde{\Delta}=\frac{\Delta}{k_F}, \quad
\Lambda(y) = -i\tilde{\lambda} \partial_y,
\end{split}
\end{align}
where $\sigma_{s} = +1 (-1)$ for $s=\uparrow (\downarrow)$.
The Hamiltonian preserves chiral symmetry:
\begin{align}
\Gamma H_{s}(k_x,y) \Gamma^{-1} = -H_{s}(k_x,y),\quad  \Gamma=\left[\begin{array}{cc}0&-i\\i&0 \end{array}\right].
\end{align}
For later convenience, we transform the Hamiltonian into the chiral basis:
\begin{align}
\begin{split}
&U_{\Gamma} H_{s}(k_x,y) U^{\dagger}_{\Gamma} = Q_{s}(k_x,y) = Q^{d_{xy}}_{s}(k_x,y) + Q^{\lambda}_{s}(y),\\
&U_{\Gamma}=\frac{1}{\sqrt{2}}\left[\begin{array}{cc}1&-i\\i&-1 \end{array}\right],\\
&Q^{d_{xy}}_{s}(k_x,y)=\left[\begin{array}{cc} 0 & -i \xi(k_x,y) - \sigma_s \Delta(k_x,y) \\ i \xi(k_x,y) - \sigma_s \Delta(k_x,y) & 0 \end{array}\right],\\
&Q^{\lambda}_{s}(y)= \left[\begin{array}{cc} 0 & -i\sigma_s\Lambda(y) \\ i\sigma_s\Lambda(y) & 0 \end{array}\right].
\end{split}
\end{align}

We first examine zero-energy states in the absence of the SOC (i.e., $\lambda=0$):
\begin{align}
\begin{split}
& Q^{d_{xy}}_{s}(k_x,y)\left[\begin{array}{cc} \phi_{s,+}(k_x,y) \\ 0 \end{array}\right]=0,\\
& Q^{d_{xy}}_{s}(k_x,y)\left[\begin{array}{cc} 0 \\ \phi_{s,-}(k_x,y) \end{array}\right]=0,
\end{split}
\end{align}
where $\phi_{s,\pm}(k_x,y)$ satisfies:
\begin{align}
\left\{ \pm i \xi(k_x,y) - \sigma_s \Delta(k_x,y) \right\} \phi_{s,\pm}(k_x,y) =0.
\end{align}
A general solution takes the form:
\begin{align}
\begin{split}
&\phi_{s,\pm}(k_x,y) = (a e^{iqy} + b e^{-iqy}) e^{\pm \sigma_s \tilde{\kappa} y}, \\
&q = \sqrt{\tilde{k}_F^2-\tilde{\kappa}^2},\quad
\tilde{k}_F=\frac{\sqrt{2m \tilde{\mu}}}{\hbar}, \quad
\tilde{\kappa}=\frac{m\tilde{\Delta}}{\hbar^2 k_F}.
\end{split}
\end{align}
Applying the boundary condition $\phi_{s,\pm}(k_x,0)=0$, we obtain:
\begin{align}
\phi_{s,\pm}(k_x,y) =c \sin (q y) e^{\pm \sigma_s \tilde{\kappa} y}.
\end{align}
Moreover, the boundary condition, $\phi_{s,\pm}(k_x,W) \propto \sin (qW) = 0$, leads to
\begin{align}
q = q_n =\frac{n\pi}{W}, \qquad &n=\pm1,\pm2,\cdots.
\end{align}
From $q = \sqrt{\tilde{k}_F^2-\tilde{\kappa}^2}$, we find the momenta at which zero-energy state exist:
\begin{align}
k_x=k_n=\mathrm{sgn}[n] \sqrt{\frac{k_F^2-q_n^2}{1+\left(\frac{\Delta}{2\mu}\right)^2}}.
\end{align}
Since $k_n$ must be real, the integer number $n$ is restricted to
\begin{align}
n=\pm1,\pm2,\cdots,\pm n_c
\end{align}
where $n_c=[Wk_F/\pi]_G$ represents the number of propagating channels per spin.
From the normalization condition:
\begin{align}
\int_0^W d y \phi^{\ast}_{s,\pm}(k_x,y) \phi_{s,\pm}(k_x,y) = 1,
\end{align}
we obtain:
\begin{align}
\begin{split}
&\phi_{s,\pm}(k_n,y)=c_{k_n,s,\pm}\sin (q_n y) e^{\pm s \kappa_n y},\\
&c_{s,\pm,n}=\sqrt{\frac{2X_n}{W}e^{\mp s \kappa_n W}},\quad
X_n=\frac{\kappa_nW}{\sinh (\kappa_n W)} \left\{ 1+\left(\frac{\Delta}{2\mu}\right)^2\frac{k_n^2}{q_n^2}\right\}.
\end{split}
\end{align}
Since $\phi_{s,+}(k_n,y)$ and $\phi_{s,-}(k_n,y)$ are degenerate, we can construct alternative expressions for the zero-energy states by taking linear combinations of these functions.
Here, we construct the wave function of the zero-energy states, $\tilde{\psi}^{\pm}_{s}(k_n,y) =\sum_{a=\pm} c_{\pm,a}\phi_{s,a}(k_n,y)$, to satisfy:
\begin{align}
\int_0^W dy \left\{ \tilde{\psi}^{\pm}_{s}(k_n,y) \right\}^{\dagger}Q^{d_{xy}}_{s}(k_n+\delta k,y) \tilde{\psi}^{\pm}_{s}(k_n,y) = \pm \epsilon_{s}(k_n + \delta k) 
\end{align}
with
\begin{align}
\epsilon_{s}(k_n + \delta k) = \left\{ \begin{array}{cl}
\mathrm{sgn}[n] |v_n| \delta k + O(\delta k^2) & \text{for } |n| \in \text{odd} \\
&\\
-\mathrm{sgn}[n] |v_n| \delta k + O(\delta k^2) & \text{for } |n| \in \text{even}\end{array} \right.,
\label{eq:abs_energy}
\end{align}
where $\pm \epsilon_{s}(k_x)$ denotes the energy eigenvalues of the Andreev bound states (ABSs).
Equation~(\ref{eq:abs_energy}) means that $\epsilon_{s}(k_x)$ forms a continuous spectrum with respect to $k_x$, as illustrated in Fig.~4(a) of the main text.
This is achieved by the wave function:
\begin{align}
\begin{split}
&\tilde{\psi}^{\pm}_{s}(k_n,y)=\left\{ \begin{array}{cl}
e^{\mp i\frac{\pi}{4}}\tilde{\varphi}^{\pm}_{s}(k_n,y) & \text{for } |n| \in \text{odd} \\ e^{\mp i\frac{\pi}{4}} \tilde{\varphi}^{\mp}_{s}(k_n,y) & \text{for } |n| \in \text{even}\end{array} \right.,\\
&\tilde{\varphi}^{\pm}_{s}(k_n,y)=\frac{1}{\sqrt{2}}\left[ \begin{array}{cc} \phi_{s,+}(k_n,y) \\ \pm i \phi_{s,-}(k_n,y) \end{array}\right],
\end{split}
\end{align}
where we specifically obtain:
\begin{align}
v_n=\frac{\hbar^2k_n}{m}\left\{1+\left(\frac{\Delta}{2\mu}\right)^2\right\}X_n.
\end{align}
In the main text, we discuss the wave function in the original basis:
\begin{align}
\psi^{\pm}_{s}(k_n,y) = U^{\dagger}_{\Gamma} \tilde{\psi}^{\pm}_{s}(k_n,y).
\end{align}

To proceed with our analysis, we treat the spin-orbit coupling, $H^{\lambda}_{s}(y)$, as a perturbation and construct the low-energy effective Hamiltonian as follows:
\begin{align}
\begin{split}
&H^{\mathrm{eff}}_s(k_x)=\left[\begin{array}{cc} A_{++} & A_{+-} \\ A_{-+} & A_{--} \end{array}\right] + O(\lambda^2),\\
&A_{\eta\eta^{\prime}}=\int_0^W dy \left\{ \psi^{\eta}_{s}(k_x,y) \right\}^{\dagger}H_{s}(k_x,y)\psi^{\eta^{\prime}}_{s}(k_x,y),
\end{split}
\end{align}
where $\psi^{\pm}_{s}(k_x,y)$ represents the wave function of the ABSs at $\lambda=0$, and $A_{\eta\eta}=\eta \epsilon_{s}(k_x)$.
In the vicinity of $k_x=k_n$, we obtain:
\begin{align}
H^{\mathrm{eff}}_s(k_n+\delta k) =&
\int_0^W dy  \left[\begin{array}{cc}
\left\{ \tilde{\psi}^{+}_{s}(k_n,y) \right\}^{\dagger}Q_{s}(k_n+\delta k,y) \tilde{\psi}^{+}_{s}(k_n,y)&
\left\{ \tilde{\psi}^{+}_{s}(k_n,y) \right\}^{\dagger}Q_{s}(k_n+\delta k,y) \tilde{\psi}^{-}_{s}(k_n,y)\\
\left\{ \tilde{\psi}^{-}_{s}(k_n,y) \right\}^{\dagger}Q_{s}(k_n+\delta k,y) \tilde{\psi}^{+}_{s}(k_n,y)&
\left\{ \tilde{\psi}^{-}_{s}(k_n,y) \right\}^{\dagger}Q_{s}(k_n+\delta k,y) \tilde{\psi}^{-}_{s}(k_n,y) \end{array}\right] + O(\delta k^2) \nonumber\\
=&H^{\mathrm{eff}}_{n,s}(\delta k) + O(\delta k^2),
\end{align}
with
\begin{align}
\begin{split}
&H^{\mathrm{eff}}_{n,s}(\delta k)=\left\{ \begin{array}{cl}
\left[\begin{array}{cc} v_n \delta k & m_n \\ m_n & -v_n \delta k \end{array}\right] & \text{for } |n| \in \text{odd} \\
&\\
\left[\begin{array}{cc} -v_n \delta k & -m_n \\ -m_n & v_n \delta k \end{array}\right] & \text{for } |n| \in \text{even}\end{array} \right.,\\
& m_n= \tilde{\lambda} \kappa_n X_n,
\end{split}
\end{align}
which is equivalent to Eq.~(13) in the main text.

\subsection{Winding number}
Next, we compute the one-dimensional winding number of the effective Hamiltonian.
By using an appropriate expression of $\psi^{\pm}_{s}(k_x,y)$, we can always construct the effective Hamiltonian preserving chiral symmetry as:
\begin{align}
\gamma H^{\mathrm{eff}}_s(k_x) \gamma^{-1}=-H^{\mathrm{eff}}_s(k_x), \quad \gamma=\left[\begin{array}{cc}0&-i\\i&0 \end{array}\right],
\end{align}
which allows us to define the winding number:
\begin{align}
N_{\mathrm{1D}}^{\mathrm{eff}} = \frac{i}{4 \pi} \sum_{s=\uparrow,\downarrow} \int dk_x \mathrm{Tr}[\gamma \{ H^{\mathrm{eff}}_s(k_x)\}^{-1} \partial_{k_x}  H^{\mathrm{eff}}_s(k_x) ].
\end{align}
The effective Hamiltonian can be generally deformed into the chiral basis as:
\begin{align}
\begin{split}
&U_{\gamma} H^{\mathrm{eff}}_s(k_x) U^{\dagger}_{\gamma} = \left[\begin{array}{cc}0&q_s(k_x)\\q^{\ast}_s(k_x)&0 \end{array}\right],\\
&U_{\gamma}=\frac{1}{\sqrt{2}}\left[\begin{array}{cc}1&-i\\i&-1 \end{array}\right],
\end{split}
\end{align}
where
\begin{align}
q_s(k_n+\delta k) =  \left\{ \begin{array}{cl}
-i v_n \delta k -m_n & \text{for } |n| \in \text{odd} \\
&\\
i v_n\delta k+ m_n & \text{for } |n| \in \text{even}\end{array} \right..
\end{align}
As shown in Ref.~[\onlinecite{sato_11}], the winding number can be further simplified to:
\begin{align}
\begin{split}
&N_{\mathrm{1D}}^{\mathrm{eff}} = \frac{1}{2} \sum_{s=\uparrow,\downarrow} \sum_{C(k_x)=0} \mathrm{sgn}[\partial_{k_x}C(k_x)] \mathrm{sgn}[R(k_x)],\\
&R(k_x) = \mathrm{Re}[q_s(k_x)], \quad C(k_x) = \mathrm{Im}[q_s(k_x)],
\end{split}
\end{align}
where the summation $\sum_{C(k_x)=0}$ is taken for $k_x$ satisfying $C(k_x)=0$.
Using this formula, we eventually obtain
\begin{align}
N_{\mathrm{1D}}^{\mathrm{eff}} = \frac{1}{2} \sum_{s=\uparrow,\downarrow}  \sum_{n=-n_c}^{n_c} \mathrm{sgn}[v_n] \mathrm{sgn}[m_n] =N_c,
\end{align}
which is also shown in the main text.

\end{document}